\documentclass[fleqn,twoside]{article}
\usepackage{espcrc2}
\usepackage{graphicx}
\usepackage[figuresright]{rotating}

\newcommand{\AmS}{{\protect\the\textfont2
    A\kern-.1667em\lower.5ex\hbox{M}\kern-.125emS}}
\def\si{{}^1\kern-.14em S_0}
\def\siii{{}^3\kern-.14em S_1}
\def\piii{{}^3\kern-.14em P_1}
\def\diii{{}^3\kern-.14em D_1}

\title{Baryons in Partially-Quenched Chiral Perturbation Theory}

\author{M. J. Savage \address{Department of Physics, University of Washington,
    Seattle, WA 98195-1560, USA}}
       
\begin{document}

\begin{abstract}
I discuss the inclusion of baryons into partially-quenched 
chiral perturbation theory  and 
describe one-loop calculations that have been performed.
\vspace{1pc}
\end{abstract}
\maketitle

%%%%%%%%%
\section{Introduction}
The motivation for including baryons into partially-quenched QCD (PQQCD), 
or
more specifically partially-quenched chiral perturbation theory (PQ$\chi$PT), 
is clear.
One wishes to take partially-quenched (PQ)
lattice simulations of observables of interest
that in the foreseeable future will be 
performed with quark
masses that are heavier than those of nature, and make rigorous predictions
for these observables in QCD.
The way that such predictions will be made is to use PQ lattice simulations to
``match'' onto a PQ effective field theory (PQEFT), and then use the explicit
quark mass dependence of the PQEFT to make predictions at the physical values
of the light quark masses, i.e. the coefficients of the local operators in the
PQEFT will be determined from the PQ lattice simulations.

%%%%%%%%%%%%%%%
\subsection{Quenching and Partial-Quenching}

Ideally, all lattice simulation would be unquenched and the
quarks would have their physical masses.  
At some point in the future this will
happen, but at this point in time and for the foreseeable future such
simulations will not be possible, simply due to the lack of 
computing power.
It is the quark loops that are disconnected from external sources (the quark
determinant), 
participating through gluons alone, that are the most time consuming to 
simulate as they  scale as a relatively high power of the inverse quark
mass.
In contrast, simulation of 
quark loops that are coupled to the external sources 
is much less time consuming.
In quenched QCD (QQCD)~\cite{Sharpe90,S92,BG92,LS96}, 
the quark determinant is set equal to unity, and thus
the time-expensive
disconnected quark loops are discarded, allowing for relatively rapid
simulation at relatively light quark masses.
From a formal standpoint, the quenched simulations correspond to introducing
an equal mass  
``ghost'' fermion for each light quark, so that 
disconnected  loop contributions
from the light quarks and the ghosts exactly cancel. Thus the 
$SU(2)_L\otimes SU(2)_R$ chiral symmetry of QCD is extended to 
$SU(2|2)_L\otimes SU(2|2)_R \times  U(1)$.
There is no formal limit in which quenched simulations can be used to make
rigorous predictions about QCD~\footnote{However, it has recently been noted
  that QQCD in the large-$N_C$ limit is identical to QCD in the 
  large-$N_C$ limit~\cite{ChenN}.}.
The discovery of PQQCD~\cite{Pqqcd,SS01} was a major step forward 
in this regard.
In addition to the ghosts that are added in QQCD, additional quarks are added 
to partially unquench the simulation, increasing the flavor symmetry of the 
theory to $SU(4|2)_L\otimes SU(4|2)_R \otimes  U(1)$.  
These additional ``sea''-quarks
participate in the disconnected loop diagrams, but if
they are heavier than the ``valence''-quarks they require
less time to simulate.  
In the limit that the mass of the sea-quarks become
equal to those of the valence-quarks (and ghosts) the matrix elements of PQQCD
between physical states
become equal to those of QCD.  
Thus one can make QCD predictions from PQQCD.

In order to describe the low-momentum dynamics 
and the light-quark mass dependence of the low-lying hadrons in QCD one,
of course, uses an EFT, such as $\chi$PT in the meson sector and heavy baryon
$\chi$PT (HB$\chi$PT) in the single baryon sector.
In extending the theory to PQQCD one has to extend 
$\chi$PT  to PQ$\chi$PT.
This allow one to determine the contributions that are non-analytic in
external momentum and $m_q$, and also to parameterize
short-distance physics in the local operators that enter order-by-order in the 
expansion.
As it is presently 
quite difficult to isolate the non-analytic contributions from the
analytic ones in lattice simulations
due to the relatively large values of $m_q$ that can
be simulated, knowledge of the low-energy EFT is vital.
The lattice simulations are used to determine the values of counterterms
that appear at any given order in the EFT.
Thus, PQ simulations will be used to fix counterterms in PQ$\chi$PT, which 
can then be used, to make QCD predictions as the $m_q$-dependence is explicit.
Such determinations are beginning to be performed in the meson 
sector~\cite{Pqqcd,SS01} .

%%%%%%%%%%
\section{Baryons in PQ$\chi$PT}

Baryon properties in quenched $\chi$PT (Q$\chi$PT) have been studied 
somewhat during the past six years, starting with the seminal paper by Labrenz
and Sharpe~\cite{LS96}.  
The representations of the baryons are  extended from
those of ordinary lie-groups to those of graded lie-groups,
$SU(3)$ to $SU(3|3)$ for three-flavor flavor QCD and from $SU(2)$ to $SU(2|2)$
for two-flavor QCD, so that the tools developed for HB$\chi$PT can be used
directly in PQHB$\chi$PT.  
In the quenched theory, the octet-baryon masses~\cite{LS96}, 
magnetic moments~\cite{S01a} and the matrix elements 
of isovector twist-2 operators~\cite{CS01a} 
have been studied at the one-loop level in the chiral expansion.

Including baryons in PQ$\chi$PT is similar to including them 
into Q$\chi$PT~\cite{CS01b,BS02b}.
One constructs the irreducible representations of the graded lie-group, 
which for two-flavors is $SU(4|2)$, 
and assigns baryon fields, just as one does  in QCD and QQCD.  
The {\bf 70} dimensional representation of baryons that contains the nucleons 
is described by a tensor of the form ${\cal B}_{ijk}$,
where $i,j,k$ run from $1$ to $6$ (we have dropped the Dirac index).
To determine the particle assignment 
it is particularly simple to use the interpolating fields 
discussed by Labrenz and Sharpe,
\begin{eqnarray}
\left[Q_i^{\alpha,a} Q_j^{\beta,b} Q_k^{\gamma,c}
-Q_i^{\alpha,a} Q_j^{\gamma,c} Q_k^{\beta,b}\right]
\epsilon_{abc} \left(C\gamma_5\right)_{\alpha\beta}
\label{eq:octinter}
\end{eqnarray}
for the representation containing the nucleons.
The objects $Q_i$ are quark field operators that include the valence, sea
and ghost quarks.
$C$ is the charge conjugation operator,
$a,b,c$ are color indices and $\alpha,\beta,\gamma$ are Dirac indices.
The interpolating field for the {\bf 44} dimensional 
representation that contains the $3/2$-baryon resonances,
${\cal T}_{ijk}$ (we have dropped lorentz and Dirac indices), 
is 
\begin{eqnarray}
\left[
Q^{\alpha,a}_i Q^{\beta,b}_j Q^{\gamma,c}_k +
{\rm cyclic}
\right]
\varepsilon_{abc} (C\gamma^\mu)_{\beta\gamma}
\ \ \ .
\label{eq:tdef}
\end{eqnarray}
One finds that
under the interchange of flavor indices~\cite{LS96},
\begin{eqnarray}
&& {\cal B}_{ijk} =  (-)^{1+\eta_j \eta_k}\  {\cal B}_{ikj}
\\
&& {\cal B}_{ijk} \ +\  (-)^{1+\eta_i \eta_j}\ {\cal B}_{jik}
\ +\ 
\nonumber\\
&& \qquad (-)^{1 + \eta_i\eta_j + \eta_j\eta_k + \eta_k\eta_i}\ 
{\cal B}_{kji}\ =\ 0
\nonumber\\
&& {\cal T}_{ijk} =
(-)^{1+\eta_i\eta_j} {\cal T}_{jik}\ =\ 
(-)^{1+\eta_j\eta_k} {\cal T}_{ikj}
\ .
\nonumber
\label{eq:bianchi}
\end{eqnarray}
with $\eta_k=+1$ for $k=1,2,3,4$ and $\eta_k=0$ for $k=5,6$.

The only baryons that appear as intermediate states in one-loop diagrams
involve either three valence quarks, or two valence quarks and
one ghost quark, or two valence quarks and one sea quark.
Thus we only need to determine the location of these baryons in the
tensors ${\cal B}$ and ${\cal T}$.
It turns out to be convenient to classify the baryons according to how they
transform under 
$SU(2)_{\rm valence}\otimes SU(2)_{\rm sea}\otimes SU(2)_{\rm ghost}$, and it
is straightforward to show that for the {\bf 70} one needs to include fields
$N_a$, $_a t$, $_a s_{bc}$, $_a\tilde t$ and $_a \tilde s_{bc}$ 
that transform as 
$(2,1,1)$, $(1,2,1)$, $(3,2,1)$, $(1,1,2)$ and $(3,1,2)$ respectively.
For the {\bf 44}, one introduces fields 
$\Delta_{abc}$, $_a x_{bc}$ and $_a\tilde x_{bc}$ that transform as 
$(4,1,1)$, $(3,2,1)$ and $(3,1,2)$ respectively.
For the three-flavor case the octet of baryons are part of a {\bf 240}
dimensional irrep of $SU(6|3)$ while the decuplet of baryon resonances are
part of a {\bf 138} dimensional irrep.

It is straightforward to construct the lagrange density describing the baryons
and their interactions with the pseudo-Goldstone bosons associated with
the spontaneous breaking of $SU(4|2)_L\otimes SU(4|2)_R$ down to $SU(4|2)_V$.
Thus the loop expansion for the low-energy properties of the nucleons in 
PQ$\chi$PT is the same as that in $\chi$PT except that the 
number of particles that can 
participate in loops is much larger, and one needs to keep track of minus
signs associated with the loops involving fermionic mesons.

\begin{table*}[ht]
\caption{One-loop calculations in PQ$\chi$PT that currently exist.}
\label{table:1}
\newcommand{\m}{\hphantom{$-$}}
\newcommand{\cc}[1]{\multicolumn{1}{c}{#1}}
\renewcommand{\tabcolsep}{2pc} % enlarge column spacing
\renewcommand{\arraystretch}{1.2} % enlarge line spacing
\begin{tabular}{@{}lll}
\hline
Observable           & \cc{$N_f=2$} & \cc{$N_f=3$} \\
\hline
Masses~\cite{CS01b,BS02b}               
& \cc{$\surd$} & \cc{$\surd$} \\
Magnetic Moments~\cite{CS01b,BS02b}      
& \cc{$\surd$} & \cc{$\surd$} \\
Matrix Elements of Twist-2 Operators~\cite{CS01b,BS02b}    
& \cc{$\surd$} & \cc{$\surd$} \\
Matrix Elements of the Axial Current~\cite{BS02b}   
& \cc{$\surd$} & \cc{$\times$} \\
$NN\pi$ Parity-Violating Interaction~\cite{BS02c}   
& \cc{$\surd$} & \cc{$\times$} \\
Nucleon Anapole Moment~\cite{BS02c}   & \cc{$\surd$} & \cc{$\times$} \\
\hline
\end{tabular}\\[2pt]
\end{table*}
Calculations of several observables at the one-loop level,
both in two- and three-flavor PQ$\chi$PT, 
have recently been completed, in addition to a tree-level analysis
of the nucleon-nucleon (NN) potential. 
A summary of what presently exists can be seen in Table~\ref{table:1}.
In the two-flavor calculations, both the valence- and sea-quarks are
non-degenerate, while in the three-flavor calculations the two light-quarks are
taken to be degenerate (both valence and sea) while the strange quark is
non-degenerate in both, i.e. $2+1$.

Instead of detailing each of these calculations, 
I wish to focus on a few aspects and 
observables that are somewhat out of the ordinary.
For further details of the mainstream observables, such as the masses,
magnetic moments and so forth I refer the reader to the appropriate papers.

%%%%%%%%%%%
\subsection{The Nucleon-Nucleon Potential}

In QCD it is well established that one pion exchange (OPE) gives the
dominant contribution to the long-distance component of the NN
(NN) potential.  OPE yields a NN potential that 
is Yukawa potential at large-distances, falling like 
$\sim e^{-m_\pi r}/r$
and also provides a $\sim 1/r^3$ short-distance tensor
contribution that mixes $S$-wave and $D$-waves.

Efforts have been made to explore the interactions between nucleons 
on the lattice by
determining their scattering lengths.  However, this is exceptionally
challenging due to the fact that the scattering lengths are unnaturally large,
$\sim 6~{\rm fm}$ in the $\siii-\diii$ coupled channels and $\sim 24~{\rm fm}$
in the $\si$ channel.
What's more the simulations have been performed in QQCD~\cite{fuku} 
and currently no unquenched or PQ calculations exist (for a recent discussion
see Ref.~\cite{Richards:2000ix}).

In both QQCD and PQQCD the long-distance component of the NN potential is very
different to that in QCD.  The hairpin interaction that gives rise to
a double pole in the iso-singlet propagator, that in QQCD
is proportional to $M_0^2$, and 
in PQQCD is proportional to $m_{vv}^2-m_{ss}^2$ and vanishes in the QCD limit,
provides the dominant contribution to the long-distance component of the NN
potential~\cite{BS02d}.
In the isospin limit, the $SU(2)$ singlet 
($\eta$, distinct from the $SU(4|2)$ singlet) 
propagator has the form
\begin{eqnarray}
G_{\eta\eta} (q^2)=
{i ( m_{ss}^2-m_{vv}^2 )\over ( q^2-m_{vv}^2 + i\epsilon )^2 }
\ \ ,
\end{eqnarray}
where
$v=u=d$ denotes valence-quarks and $s=j=l$ denotes sea-quarks.
The mass $m_{vv}$ denotes the mass of a meson containing two valence-quarks,
and
$m_{sv}$ denotes the mass of a meson containing one valence-quark and one
sea-quark.
Fourier transforming the momentum-space NN potential gives
\begin{eqnarray}
&&V^{(PQ)} (r)\ =\ 
{{\bf\sigma}_1\cdot {\bf\nabla} {\bf\sigma}_2\cdot {\bf\nabla} \over 8\pi f^2}
\ e^{-m_\pi r} 
\\
&&\qquad\left( g_A^2\ {{{\bf\tau}_1\cdot{\bf\tau}_2}\over r} -
g_0^2\ {{\left(m_{\eta_j}^2-m_\pi^2\right)}\over{2 m_\pi}}
\right) 
\ \ ,
\nonumber
\end{eqnarray}
which does not exhibit Yukawa type fall off at large distance 
but rather falls exponentially.
Therefore, unfortunately, efforts to observe 
the long-distance behavior of the NN interaction
in QQCD and PQQCD will need to subtract this large contribution.

%%%%%%%%%%%%%%%
\subsection{Charges}

When one considers processes involving electroweak gauge fields, there is an
additional freedom in charge assignments that is not present in QCD.
One requires that QCD is recovered in the limit where the mass of the 
sea-quarks become identical to the mass of the valence quarks, and thus the
electromagnetic charge matrix in QCD, 
${\cal Q}^{\rm QCD}={\rm diag}(+{2\over 3},-{1\over 3})$, is extended to 
${\cal Q}={\rm diag}(+{2\over 3},-{1\over 3},q_j,q_l,q_j,q_l)$
in PQQCD.  The charges $q_j$, $q_l$ are arbitrary, and in the QCD limit,
observables become independent of them.
However, away from the QCD limit quantities do depend upon them, and it may be
useful to make specific choices for particular observables.
For instance, choosing $q_j=q_l=0$ means that disconnected loops coupling
to a photon involve only valence-quarks, while choosing 
$q_j=+{2\over 3}$ and $q_l=-{1\over 3}$ means that 
disconnected loops coupling to a photon involve only sea-quarks.
Thus, numerically, 
there may be some advantage in choosing the latter set of charges 
over the former, but this remains to be explored.
In addition, there are sets of charges that minimize the contribution
from one-loop diagrams in PQ$\chi$PT.
As higher orders are parametrically suppressed, this will lead to the smallest
variation in the matrix element with respect to changes in $m_q$.

The non-uniqueness in the extension of electroweak operators from QCD to PQQCD
is not confined to the electric charges.  It is also present for the twist-2
operators~\cite{CS01b}, and for four-quark operators, 
such as those responsible for 
$K\rightarrow \pi\pi$~\cite{GP01a} and for $NN\pi$ 
parity-violation~\cite{BS02c}.

%%%%%%%%%%%%%%%%
\subsection{Proton Magnetic Moment}

As an example of a calculation in PQ$\chi$PT, we discuss the 
proton magnetic moment, $\mu_p$, up to order ${\cal O}(m_Q^{1/2})$
in the chiral expansion.
The leading order contributions to the nucleon magnetic moment arise from the 
dimension-5 operators
\begin{eqnarray}
& & {\cal L}^{(5)}=
{e\over 4 M_N}\ F_{\mu\nu}\  \left[ 
\mu_\alpha\ \left(\overline{\cal B}\sigma^{\mu\nu} {\cal B} {\cal Q}\right)
\right.\\ & & \left.
\ +\ 
\mu_\beta\ \left(\overline{\cal B}\sigma^{\mu\nu} {\cal Q}
{\cal B}\right)
\ +\ 
\mu_\gamma\ \left(\overline{\cal B}\sigma^{\mu\nu}
{\cal B}\right) {\rm str}{\cal Q}
\right] 
\ ,
\nonumber
\end{eqnarray}
where the brackets $\left(..\right)$ denote contractions of the flavor indices
as discussed in Ref.~\cite{LS96}, and ``${\rm str}$'' denotes a supertrace.
At tree-level, the coefficients $\mu_\alpha, \mu_\beta$ and $\mu_\gamma$ are 
related to the isoscalar, $\mu_0$, 
and isovector, $\mu_1$, magnetic moments of the nucleon,
\begin{eqnarray}
& & \mu_0 = {1\over 6}\left(\mu_\alpha+\mu_\beta+2\mu_\gamma\right)
\nonumber\\
& &  \mu_1 = {1\over 6}\left(2 \mu_\alpha-\mu_\beta\right)
\ \ .
\end{eqnarray}
Up to ${\cal O}(m_Q^{1/2})$, the magnetic moment of the proton can be written
as
\begin{equation}
\mu_p=\alpha_p + {M_N\over 4\pi f^2} \left[\beta_p+\beta_p^\prime\right]
+...
\ \ ,
\end{equation}
where
\begin{eqnarray}
& & \alpha_p=\mu_0+\mu_1
\\
& & \beta_p=-g_A^2 m_{vv}
\nonumber\\
& &\ \ + \left(m_{sv}-m_{vv}\right)
\left[{8\over 9} g_A^2 + {4\over 9} g_A g_1 + {1\over 18} g_1^2
\right]
\nonumber\\
& &\ \ +q_{jl}\left(m_{sv}-m_{vv}\right)
\left[ {2\over 3} g_A^2 + {1\over 3} g_A g_1 + {1\over 6} g_1^2\right]
\nonumber\\
& &
\beta_p^\prime={1\over 27} g_{\Delta N}^2 \left[-6 {\cal F}_{vv} + 
q_{jl} {9\over 2} \left( {\cal F}_{vv}-{\cal F}_{sv} \right)\right]
\ \ ,
\nonumber
\end{eqnarray}
with $q_{jl} = q_j+q_l$.  
The function ${\cal F}_{ij} = {\cal F}(m_{ij},\Delta,\mu)$ is given by 
\begin{eqnarray}
\pi {\cal F}=
\eta\log\left({\Delta-\eta
\over \Delta+\eta}\right)
 -\ \Delta\log\left({m^2\over\mu^2}\right)
\ ,
\label{eq:magfun}
\end{eqnarray}
where $\eta = \sqrt{\Delta^2-m^2+i\epsilon}$,
$\Delta$ is the $\Delta$-nucleon mass splitting in the chiral limit, and
$\mu$ is the dimensional regularization renormalization scale.
$g_A$ is the usual axial coupling constant, 
while $g_1$ is an additional axial
coupling that contributes in PQ$\chi$PT, but whose 
contribution must vanish in the QCD limit.
$g_{\Delta N}$ is the $\Delta N\pi$ coupling constant.

There are two distinct contributions from the loop diagrams away from the QCD
limit. One is a contribution that is independent of the choice of charges that
vanishes like $m_{sv}-m_{vv}$, while the other also vanishes like 
$m_{sv}-m_{vv}$ but depends upon the choice of charges of the sea- and 
ghost-quarks.
Therefore, one is able to find sets of charges for which the one-loop
contribution to the proton magnetic moment from intermediate states in the 
{\bf 70} vanish, however, due to the non-trivial mass-dependences arising from 
intermediate states in the {\bf 44}, the one-loop contributions cannot be
eliminated entirely.

%%%%%%%%%%%%
\subsection{$NN\pi$ Parity Violation}

While flavor-changing parity-violating (PV) interactions are well 
understood theoretically and a
great deal of precise data exists, knowledge of flavor-conserving
parity-violation is rather primitive. Flavor-conserving
parity-violation continues to be an area of intense investigation
in the nuclear physics community. 
Its study is presently serving both to uncover the structure of the
nucleon in electron-scattering experiments such as SAMPLE\cite{sample}, and to
determine PV flavor-conserving couplings between pions
and nucleons\cite{cesium,Fluorine}. 
The
problems that are encountered in this sector are both
experimental and theoretical.  On the experimental side, the 
PV signals, unlike those in flavor-changing processes, appear as small
deviations in either a strong or an electromagnetic process, such as PV in
$ep\rightarrow ep$, or in the circular polarization of the $\gamma$-ray emitted
in $^{18}F^* \rightarrow ^{18}F \gamma$~\cite{Fluorine}.  The current situation
is somewhat confused by the fact that measurements of parity-violation in 
atoms and nuclei do not give rise to a consistent set of couplings
between hadrons~\cite{Haxton:2001ay}. However, it is important to keep in 
mind that many of the
``experimental'' determinations of these couplings require theoretical inputs
with varying degrees of reliability.  Recently, it has been reemphasized that
measurements of PV observables in the single-nucleon sector would significantly
ameliorate the situation by eliminating many-body
uncertainties~\cite{BSpv,Chpv}.  Despite the inherent difficulty of such
experiments, there are ongoing efforts to measure PV processes in systems with
only one or two nucleons, such as the angular-asymmetry in 
$\vec np\rightarrow d\gamma$~\cite{snow}. 
Such measurements should provide a reliable
determination of the leading-order (LO), momentum-independent weak $\pi NN$ 
coupling constant,
$h_{\pi NN}^{(1)}$.

On the theoretical side, despite heroic efforts to
model~\cite{DDH,Henley} hadronic matrix elements of the four-quark
operators that appear in the low-energy effective theory of the standard model,
there are no reliable calculations of the PV couplings between hadrons. A first
principles calculation of $h_{\pi NN}^{(1)}$ in lattice QCD would therefore be
extremely welcome.  This would require a lattice QCD simulation of a correlator
with three hadronic sources interacting via a four-quark operator.
Unfortunately, chiral symmetry does not allow one to relate the $\pi NN$
correlator to a correlator without the pion. On the bright side, the structure
of the four-quark weak Hamiltonian requires a flavor change in the nucleon and
therefore there are no disconnected diagrams to be computed on the lattice.

The extraction of $h_{\pi NN}^{(1)}$ from $N\rightarrow N\pi$
requires an injection of energy at the PV weak vertex
which can occur because  the weak operator is inserted on one time-slice only.
Therefore, we must include contributions from operators that are total
derivatives, which usually vanish.
Recently, 
chiral perturbation theory has been used to describe $K\rightarrow \pi\pi$
with the  kinematics appropriate for a lattice determination of the matrix
elements of the relevant four-quark operators,
$m_K^{\rm latt}=m_\pi^{\rm latt}$ and $m_K^{\rm latt}=2 m_\pi^{\rm latt}$,
including the necessary total derivative terms~\cite{kpipi}.
In QCD, the LO Lagrange density describing PV interactions 
is given by
\begin{eqnarray}
& & {\cal L}_{\rm wk} \ =\  -h_{\pi NN}^{(1)}\  {f\over 4} \ 
\overline{N} \left[\ X_L^3\ -\ X_R^3\ \right] N
\\
&&\qquad \ -\ 
h_{\pi \Delta\Delta}^{(1)}\  {f\over 4} \
\overline{T}^{abc,\mu}
\left[\ X_L^3\ -\ X_R^3\ \right]^d_c T_{abd,\mu}
\nonumber\\
 & & \rightarrow  
i \pi^-  \left[h_{\pi NN}^{(1)}  \overline{n} p 
+
{h_{\pi \Delta\Delta}^{(1)}\over\sqrt{3}}
\overline{\Delta}^{+\mu}\Delta^{++}_\mu
\right.\nonumber\\&&\left.
+
{2 h_{\pi \Delta\Delta}^{(1)} \over 3}\overline{\Delta}^{0\mu}\Delta^{+}_\mu
+
{ h_{\pi \Delta\Delta}^{(1)} \over\sqrt{3}}
\overline{\Delta}^{-\mu}\Delta^{0}_\mu
\right]+ {\rm h.c.}
\ .
\nonumber
\label{eq:weaktree}
\end{eqnarray}
while the Lagrange density at NLO is (keeping only nucleon operators)
\begin{eqnarray}
{\cal L}_{\rm wk}^{(NLO)}= 
{h_{D}^{(1)}\over 4} \ 
i v\cdot D\ \overline{N} \left[\ X_L^3\ -\ X_R^3\ \right] N
\ \ \ ,
\label{eq:NLOweak}
\end{eqnarray}
where $v^\mu$ is the nucleon four-velocity.  This is the leading contribution
from a heavy baryon reduction of 
$i D^\mu \overline{N} \gamma_\mu \left[\ X_L^3\ -\ X_R^3\ \right] N$.
Given baryon number conservation, the total derivative gives a non-zero
contribution from the energy and momentum injected by the 
$X_L^3\ -\ X_R^3$ insertion.
Working in the frame where the initial state nucleon (proton) is at rest,
$v^\mu = (1,0,0,0)$,
the amplitude at NLO resulting from eq.~(\ref{eq:weaktree}) and 
eq.~(\ref{eq:NLOweak}) is 
\begin{eqnarray}
&&{\cal A}_{\overline{n}p\pi}\ =\ 
\langle n\pi | i \int d^3 {\bf x}\  {\cal L}^{\Delta I=1} (E) 
| p\rangle
\nonumber\\
&& = -\overline{U}_n \ 
\left[\ h_{\pi NN}^{(1)}\ +\ h_{D}^{(1)} {E\over f}\ \right]
U_p
\ \ \ .
\label{eq:NLOtree}
\end{eqnarray}
where $E$ is the energy injected by the weak vertex.
In order to produce an on-shell $n\pi$ final state, the injected energy must
exceed $E\ge m_\pi + M_n-M_p$.
Near threshold, where the final state neutron and pion are at rest and 
$E =  m_\pi + M_n-M_p$, 
the contribution from the total-derivative operator, $h_{D}^{(1)}$,
scales as $\sim m_q^{1/2}$, and is formally dominant over
loop corrections and counterterms~\cite{BS02c}.

%%%%%%%%%%%%%%%
\section{Conclusions}

Lattice simulations will first make rigorous predictions about observables
by matching onto the appropriate EFT  and using its explicit
quark mass dependence to extrapolate from the lattice quark masses to those of
nature.
Therefore the EFT's must be known, which is the case in the meson sector and
single baryon sector.  However more work is required in the multi-nucleon
sector to be sure that the candidate theory~\cite{NNEFT}
(for a review see Ref.~\cite{Be00}), in fact, is consistent and 
converges.
PQ simulations represent the future of this field until fully
unquenched calculations can be performed at the physical values of the 
light quark masses, and in this work we have 
made the small step of including baryons in PQQCD.
We have constructed PQ$\chi$PT and shown that there are some interesting
features beyond QCD.
This field is just beginning and calculations beyond one-loop level are
certainly required in order to understand the convergence properties 
of PQ$\chi$PT
and the uncertainties introduced in chiral extrapolations.

\vskip 0.2in

I am indebted to my colleagues, Silas Beane and Jiunn-Wei Chen, 
my collaborators on these works.

\end{document}